\documentclass[sigconf]{acm-contents/acmart}
\AtBeginDocument{%
  }

\copyrightyear{2025}
\acmYear{2025}
\setcopyright{rightsretained}
\acmConference[CHI EA '25]{Extended Abstracts of the CHI Conference on
Human Factors in Computing Systems}{April 26-May 1, 2025}{Yokohama, Japan}
\acmBooktitle{Extended Abstracts of the CHI Conference on Human Factors in
Computing Systems (CHI EA '25), April 26-May 1, 2025, Yokohama,
Japan}\acmDOI{10.1145/3706599.3719964}
\acmISBN{979-8-4007-1395-8/2025/04}

\usepackage{graphicx}
\usepackage{subfig}
\usepackage{enumitem}

\begin{document}

\title{Fewer Than 1\% of Explainable AI Papers Validate Explainability with Humans}

\author{Ashley Suh}
\email{ashley.suh@ll.mit.edu}
\orcid{0000-0001-6513-8447}
\affiliation{%
  \institution{MIT Lincoln Laboratory}
  \city{Lexington}
  \state{MA}
  \country{USA}
}

\author{Isabelle Hurley}
\email{isabelle.hurley@ll.mit.edu}
\orcid{0009-0006-4031-9018}
\affiliation{%
  \institution{MIT Lincoln Laboratory}
  \city{Lexington}
  \state{MA}
  \country{USA}
}

\author{Nora Smith}
\email{nora.smith@ll.mit.edu}
\orcid{0000-0001-5255-9226}
\affiliation{%
  \institution{MIT Lincoln Laboratory}
  \city{Lexington}
  \state{MA}
  \country{USA}
}

\author{Ho Chit Siu}
\email{hochit.siu@ll.mit.edu}
\orcid{0000-0003-3451-8046}
\affiliation{%
  \institution{MIT Lincoln Laboratory}
  \city{Lexington}
  \state{MA}
  \country{USA}
}
\renewcommand{\shortauthors}{Suh et al.}

\begin{abstract}
This late-breaking work presents a large-scale analysis of explainable AI (XAI) literature to evaluate claims of human explainability. We collaborated with a professional librarian to identify 18,254 papers containing keywords related to explainability and interpretability. Of these, we find that only 253 papers included terms suggesting human involvement in evaluating an XAI technique, and just 128 of those conducted some form of a human study. In other words, fewer than 1\% of XAI papers (0.7\%) provide empirical evidence of human explainability when compared to the broader body of XAI literature. Our findings underscore a critical gap between claims of human explainability and evidence-based validation, raising concerns about the rigor of XAI research. We call for increased emphasis on human evaluations in XAI studies and provide our literature search methodology to enable both reproducibility and further investigation into this widespread issue.
\end{abstract}




\maketitle

\section{Introduction}

The growing accessibility of artificial intelligence (AI) promises significant benefits for our society~\cite{dash2019application, rolnick2022tackling, ly2017fully}, as well as a number of critical challenges~\cite{board2019ai, cath2018governing, borenstein2021emerging}. 
A pressing obstacle in the field is the lack of explainability, trustworthiness, and transparency in AI technologies~\cite{rudin2019stop, saeed2023explainable, rudin2019we, holzinger2017we}. 
With AI systems being deployed to diagnose medical conditions, navigate vehicles on the road, and forecast natural disasters, understanding a model's behavior, recommendations, and reasoning has become imperative~\cite{edwards2021eu}. 
Yet, habitually used yet nondescript metrics like ``accuracy'' and ``confidence'' fail to capture the whole story of an AI model nor how it will be interpreted by humans in real-world contexts~\cite{suh2024metrics}.

The potential for skepticism and hesitation in trusting AI has driven the community to develop techniques in interpretable and explainable AI (XAI), concepts that are generally interchangeably used~\cite{lipton2018mythos}, along with transparent and trustworthy AI\footnote{While a precise definition for interpretability is not universal, Doshi-Velez and Kim define it as \textit{the ability to explain or present AI/ML systems in understandable terms to a human}~\cite{doshi2017towards}. We refer to these methods collectively as research in explainable AI (XAI).}. Common XAI approaches include rule-based reasoning, visualizations of deep learning processes, text-based explanations, and example-driven counterfactuals~\cite{saeed2023explainable}. 
Altogether, the developers of these explainability methods argue that AI models are now more usable, interpretable, and effective for human end-users than their non-explainable ``black box'' counterparts.  

However, research has repeatedly shown that explainability methods are often never tested with actual end-users~\cite{wells2021explainable, nauta2023anecdotal, miller2017explainable} -- leading to mismatched user expectations, unused models, and diminished trust in AI~\cite{koo2015did, algorithmia2021enterprise, siu2021evaluation, buccinca2021trust}. A small survey conducted by Miller et al.~\cite{miller2017explainable} revealed that, from 23 papers published at the \textit{2017 Explainable AI workshop at the International Joint Conference on Artificial Intelligence}, only 4 articles referenced relevant social science research, only 1 of them built a model on this research, and none of the 23 included any empirical testing. Wells and Bednarz’s~\cite{wells2021explainable} found that 17 out of the 25 papers they surveyed on explainable AI for reinforcement learning did not include any user testing. Similarly, Nauta et al.~\cite{nauta2023anecdotal} surveyed 300 papers published at top AI/ML conferences and found that 1 in 3 papers evaluated their methodology exclusively with anecdotal evidence, and only 1 in 5 papers evaluated with actual users. 

Our recurrent observation of explainability claims that lack human evidence motivated this late-breaking work. We sought to understand precisely how many papers across all XAI literature are promoting explainable, trustworthy, transparent, etc. methods without empirical testing. While there are many surveys of the XAI literature~\cite{saeed2023explainable, wells2021explainable}, what they tend to focus on is questions about what methods the field is using for explainability, what metrics or quantitative experiments are being carried out, or in some cases, what the state of the empirical testing is like. Furthermore, Miller et al.'s 2017 survey~\cite{miller2017explainable} -- while distinct from these surveys -- was self-described as a ``light scan of the literature,'' only reviewing 23 papers. Similarly, Wells and Bednarz’s~\cite{wells2021explainable} reviewed 25 papers, while Nauta et al.~\cite{nauta2023anecdotal} reviewed 300.


We build on these related works by significantly expanding the scope of reviewed papers to include all relevant XAI literature and preprints in the Scopus database~\cite{burnham2006scopus}. We collaborated with a professional librarian who helped us gather \textbf{over 18,000} XAI papers and filter them based on claims about human explainability. The research question we aimed to answer was, \textit{``Of all the XAI papers that claim human explainability, how many validate those claims with empirical evidence?''}
Shockingly, we found that only \textbf{0.7\%} of XAI papers in the last decade test their methods with humans evaluating the explanations. Based on our literature review, the field is nearly entirely devoid of empirical evidence about its central claims of human explainability. For the remainder of the paper, we outline our methodology (which includes our exact Scopus database search for others to reproduce our work), the results of our literature survey, and a call to action for the broader community.

\section{Methods}
\begin{table*}[t]
\centering
\caption{Keyword search for gathering relevant literature using the Scopus database. The left column describes the high-level search strategy in natural language. The middle column shows the keyword search for scraping all XAI-related papers (e.g., papers that include terms related to ``explainability'', ``interpretability'', ``XAI'', etc.). The last column scrapes the same set of XAI papers but adds constraints related to human end-user testing (e.g., papers including ``human-subject'', ``end-user'', ``experiment'', ``feedback'', ``interview'', and so on). } 
\renewcommand{\arraystretch}{1.3}
\centering
\begin{tabular}{l|l|l}
\toprule
\textbf{Scopus Search Strategy} 
& \textbf{All XAI papers}                                                               
& \textbf{XAI with claims about humans} \\
\midrule
XAI papers
& \begin{tabular}[c]{@{}l@{}}
TITLE-ABS(((explainab* OR\\
understandab* OR ``interpretab*''\\
OR transparen*) W/1 (``artificial \\
intelligence'' OR ``AI'' OR\\
``machine learning'' OR ``ML'')) OR XAI)
\end{tabular} 

& \begin{tabular}[c]{@{}l@{}}
TITLE-ABS(((explainab* OR\\
understandab* OR ``interpretab*''\\
OR transparen*) W/1 (``artificial \\
intelligence'' OR ``AI'' OR\\
``machine learning'' OR ``ML'')) OR XAI)
\end{tabular}  \\
\begin{tabular}[c]{@{}l@{}} AND\\ 
Human end-user testing
\end{tabular} 
& 
& \begin{tabular}[c]{@{}l@{}}AND\\ 
TITLE-ABS((human OR ``human-subject''\\
OR user OR ``end-user'') Pre/1 (experiment*\\
OR test OR trial OR evaluat* OR validat*\\
OR feedback OR interview))
\end{tabular}  \\
\begin{tabular}[c]{@{}l@{}}AND\\ 
English
\end{tabular}                  
& \begin{tabular}[c]{@{}l@{}}AND\\
LANGUAGE (english)
\end{tabular}                                                                                                                                                
& \begin{tabular}[c]{@{}l@{}}AND\\ 
LANGUAGE (english)
\end{tabular}  \\
\begin{tabular}[c]{@{}l@{}}NOT\\ 
Review papers
\end{tabular}          

& \begin{tabular}[c]{@{}l@{}}AND NOT\\ 
(TITLE(review OR survey OR overview)\\
OR DOCTYPE(cr OR re))
\end{tabular}     

& \begin{tabular}[c]{@{}l@{}}AND NOT\\ 
(TITLE(review OR survey OR overview)\\
OR DOCTYPE(cr OR re)) 
\end{tabular}     
\\
\bottomrule
\end{tabular}

\label{tab:scopus-search}
\end{table*}


We worked with a professional librarian over the course of three months to refine and generate advanced searches that could identify all existing XAI literature, as well as XAI literature with human-subject testing. This process involved iterating over relevant XAI search criteria~\cite{miller2017explainable}, reviewing the results of the keyword search, and refining the keywords to include or exclude relevant or irrelevant papers. This process was repeated multiple times. 

For the literature search, we selected the Scopus database for its broad coverage and ability to include the published and preprint literature. We wanted to include preprints in our search as much of the AI field publishes their results on arXiv, OSF, or another preprint platform due to the speed at which the field develops. However, we did not include review articles, workshop papers, or non-English papers. We discuss this limitation further in Section~\ref{sec:future}.

A total of 18,254 XAI papers were identified (13,839 published papers, 4,394 preprints), we refer to this set of papers as \textbf{All XAI papers}. We then filtered to a smaller subset of 253 papers which contained terms consistent with human subjects participating in testing, feedback, validation, interviews, and so on. This subset is referred to as papers with \textbf{Claims about humans} for the remainder of this work. It is important to note that papers that fall into the `claims about humans' category are papers that contain specific terms like `human feedback,' `end-user evaluation,' `user interview,' and so on -- see the right column in Table~\ref{tab:scopus-search}. We provide the full keyword search for anyone that would like to reproduce or refine our same literature review. 

\subsection{Scoring Criteria}
The final subset of papers was manually reviewed by the authors.  
We use the \textbf{validation} criteria from Miller et al.~\cite{miller2017explainable} and give each paper in review (i.e., papers in the `Claims about humans' category) one of two scores: 1 when the paper itself validates explainability claims with humans, and 0 otherwise. In the future we plan to code the broader set of XAI papers using Miller et al.’s \textbf{data-driven} criteria, which scores a paper based on whether it cites or implements work from the social sciences. However, we note that Miller et al.’s data-driven criteria -- which was determined in 2017 -- may be outdated. We discuss further in Section~\ref{sec:future}. 
For now, we limit our scope to only the use of humans in a behavioral study to evaluate the XAI method (i.e., the validation criteria) for simplicity.  

Each article in the final review set was scored by 3 people in total with XAI backgrounds. To assign a validation score, we checked if any human experiments were conducted to test the proposed XAI method. Disagreements in scoring were broken collaboratively.  

We observed several reasons to assign a 0 in our validation scoring. They include:   

\begin{itemize}[topsep=1pt, partopsep=0pt,itemsep=2pt,parsep=0pt]
    \item The paper claims it is explainable-by-design, so it does not require human experiments. 
    \item The paper contributes a framework or evaluation method that claims to be explainable but does not test the framework or evaluation method with humans.
    \item The paper claims human explainability because it used data derived from humans in it its training of the AI model or XAI method but does not evaluate with humans. We discuss this further in Section~\ref{sec:no-humans}.
    \item The paper conducted a quantitative study comparing to human-annotated data or human labels (e.g., as a baseline), but does not provide human validation.
\end{itemize}

Finally, we excluded 16 papers from the final set that were agreed upon as off topic. Off topic papers did not present an XAI method but rather were meta-reviews of the field of XAI, philosophical arguments about explainability, or were summaries of several papers by particular graduate students. 
\section{Results}
\label{sec:results}

\begin{table*}
\centering
\caption{Results of scoring 253 explainable AI papers that contained keywords consistent with humans participating in testing, providing feedback for, etc. an XAI technique or system (see Table~\ref{tab:scopus-search}). Our validation criteria follows Miller et al.'s~\cite{miller2017explainable}, where papers receive a `0' if they did not validate with humans and a `1' if they did.}
\renewcommand{\arraystretch}{1.3}
\begin{tabular}{llll}
\toprule
\textbf{Criteria} & \textbf{0 (No human evaluation)} & \textbf{1 (Yes human evaluation)} & \textbf{Off topic}\\
\midrule 
Validation  & 109                              & 128                               & 16\\
\bottomrule
\end{tabular} 
\label{tab:validation-count}
\end{table*}

In total we scored 237 XAI papers that were both on topic and had terms consistent with claims about human explainability. The breakdown of our scores is shown in Table~\ref{tab:validation-count}.
From the 237 papers, the difference between those that actually conduct a human evaluation and those that do not is very small (54\% of papers provide an evaluation, 46\% do not). If you consider this number against the larger body of explainable AI work, it implies only 0.7\% of XAI papers provide evidence from a human evaluation. Not only does this mean that nearly half of the XAI papers that directly claim human explainability never test this claim with humans, but that for every 1,000 XAI papers that do not explicitly mention humans, fewer than 8 of them provide explainability validation with empirical evidence. 

\begin{figure*}
    \subfloat[Comparison of all XAI literature we reviewed.]{%
        \includegraphics[width=.49\linewidth]{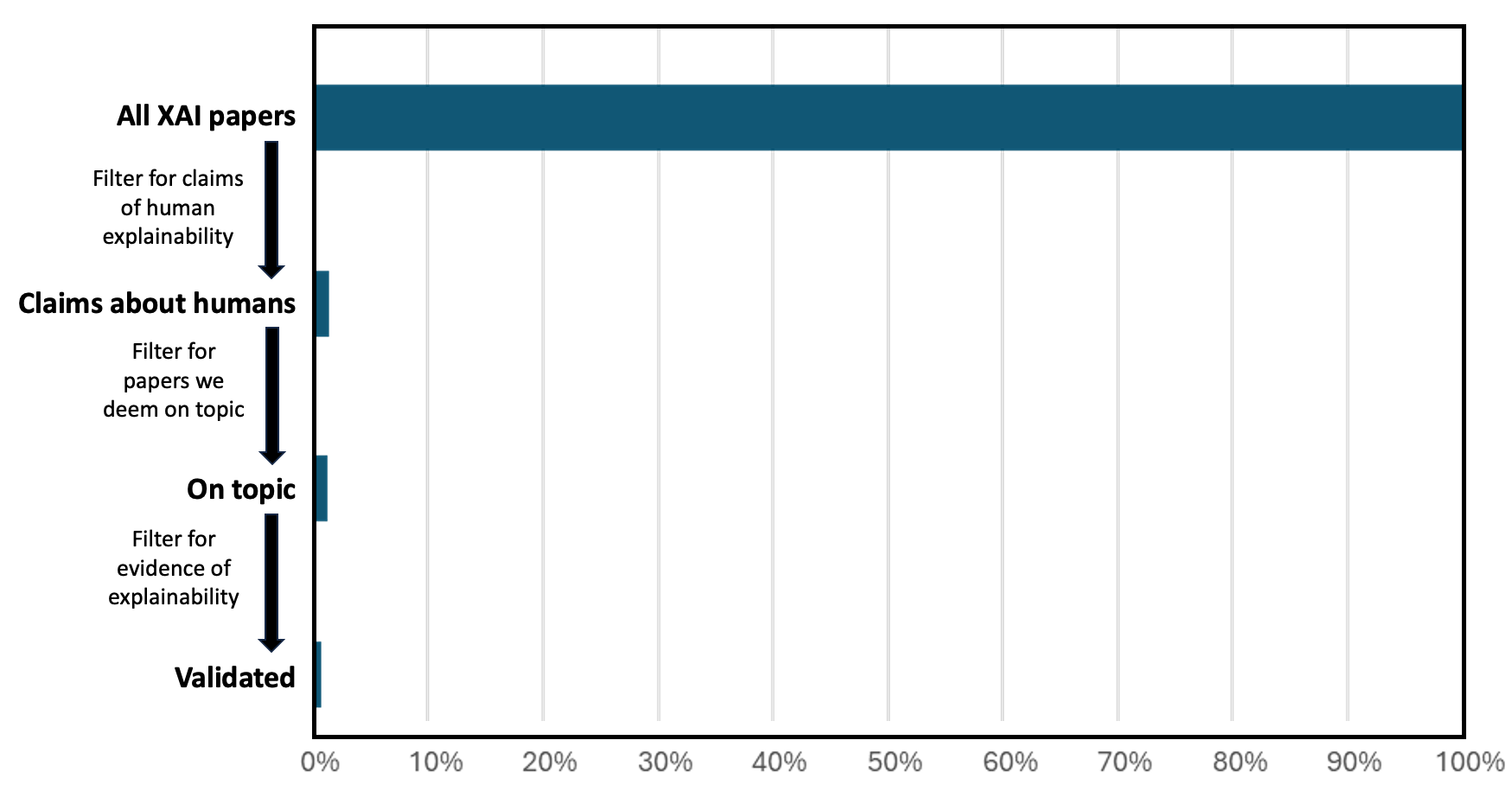}%
        \label{subfig:xai-percent}%
    }\hfill
    \subfloat[Zoomed in comparison of filtered XAI papers.]{%
        \includegraphics[width=.49\linewidth]{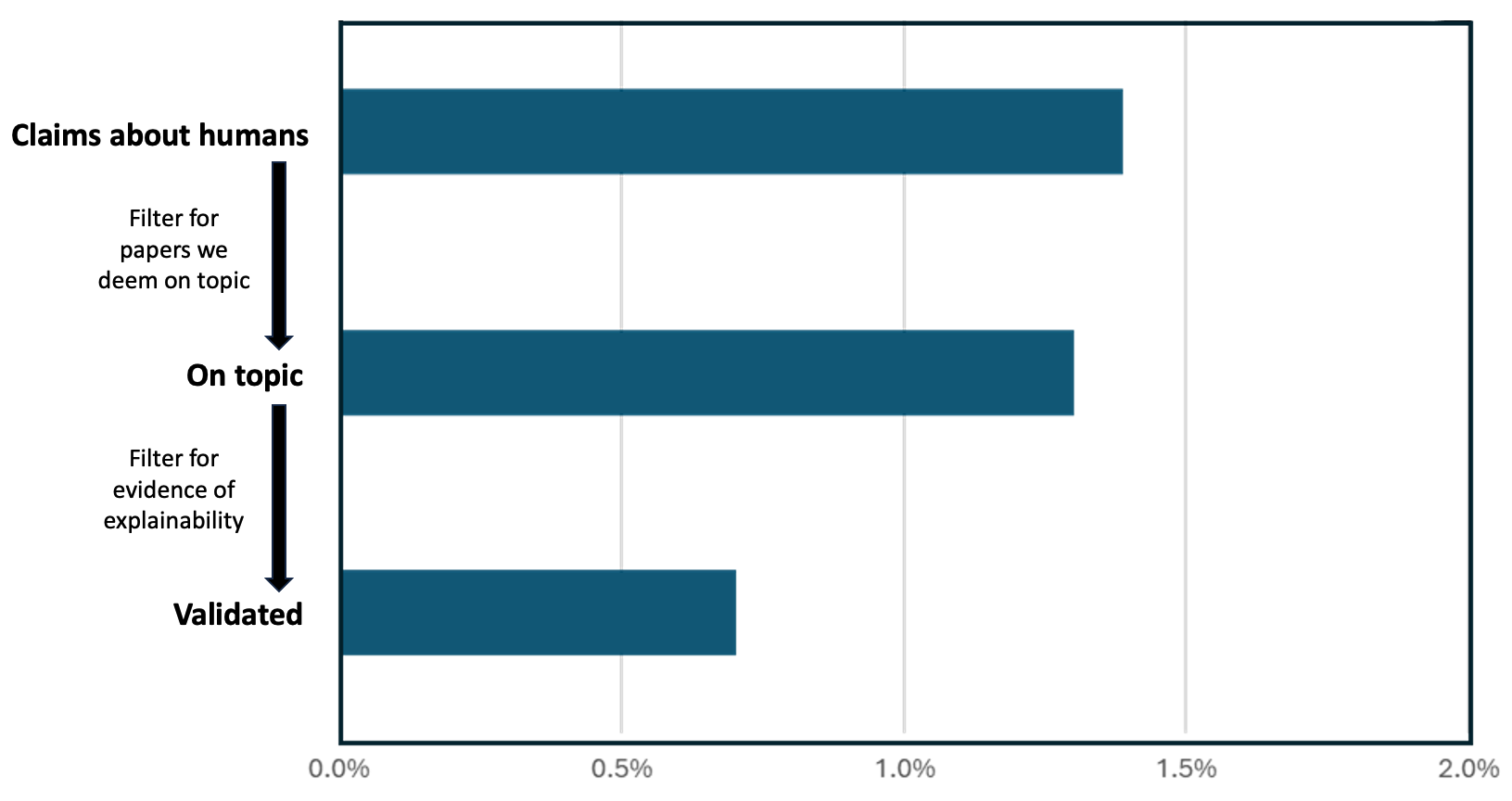}%
        \label{subfig:xai-percent-zoomed}%
    }
    \caption{Distribution comparing all XAI literature from our Scopus search (Table~\ref{tab:scopus-search}), including those we scored. `All XAI papers' is the superset that encompasses all papers with keywords related to explainability, interpretability, etc. `Claims about humans' papers are a subset of those that were filtered based on keywords related to human explainability. `On topic' papers are a subset of those that we filtered to exclude meta reviews, surveys, etc. `Validated' papers are a subset of those that provided empirical evidence of explainability. On the right a zoomed in version of the last three bars from the left figure are shown, note the change in x-axes.}
    \Description{Two subfigures depicting bar charts side by side. The first subfigure on the left shows a bar that represents 100\% of the x-axis, where the x-axis ranges from 0\% and 100\%. The bottom three bars below the first bar are extremely small in comparison. The right subfigure shows the same bottom three bars but with a smaller x-axis from 0\% to 2\%.}
    \label{fig:xai-dist}
\end{figure*}

The distribution of our down-selection results are shown in Figure~\ref{subfig:xai-percent}, with a zoomed in version of the last three bars shown in Figure~\ref{subfig:xai-percent-zoomed} -- note the change in the x-axes. `All XAI Papers' is the entirety of XAI literature that came back from our Scopus search (Table~\ref{tab:scopus-search}, middle column), `Claims about humans' is the subset of those XAI papers that use keywords related to human-subjects, end-users experiments, tests, trials, etc. (Table~\ref{tab:scopus-search}, right column), `On topic' is a smaller subset of those papers that we deemed on topic (were not meta-reviews of the literature, survey papers, etc.), and `Validated' is the smallest subset of those papers that actually provided empirical evidence of human explainability. The same data is presented in Table~\ref{tab:xai-dist}, where each row is a subset of the one above, corresponding to the previously described search and review refinements.

\begin{table*}
\centering
\caption{Table representation of our down-selection results. Our down-selection criteria is described in Section~\ref{sec:results} and partially illustrated by keyword search in Table~\ref{tab:scopus-search}.} 
\renewcommand{\arraystretch}{1.3}
\begin{tabular}{llll}
\toprule
\textbf{Down-selection criteria} & \textbf{Count} & \textbf{\% of all XAI papers over the last decade} \\
\midrule
Has explainability-related keywords & 18,254         & 100\%                                              \\
Has claims about human explanability & 253                  & 1.39\%                                             \\
Were on-topic (not surveys, etc.)     & 237                  & 1.30\%                                             \\
Were validated on human subjects      & 128                  & 0.70\%                                             \\
\bottomrule
\end{tabular}
\label{tab:xai-dist}
\end{table*}

If we consider the inverse of the percentages above, we see that over 98\% of XAI papers do not appear with human evaluation keywords, and over 99\% of XAI papers do not contain empirical human evaluation. There is naturally some difficulty in being precise with inclusion or exclusion criteria, but the results are nonetheless striking. Even if our search criteria were dramatically incorrect, say, an order of magnitude off, a ten-fold increase in validation rates in the literature would only result in 7\% of works being tested with humans. We discuss refining our search criteria in the future in Section~\ref{sec:future}.

Since our work is inspired by Miller et al.'s~\cite{miller2017explainable}, we provide a few relevant insights.
Of the 253 papers in our search, only 11 cited Miller et al. Of those 11, 6 provided a human evaluation and 5 did not, roughly matching the overall rates of human evaluation in the literature. Thus, even the papers that point to a need for such evaluations do not perform them. That said, in general, the number of citations for Miller et al. has increased since publication (roughly 15 additional citations between 2018-2019 and 100 additional citations between 2023-2024). 



\section{Discussion of Results}

\subsection{Studies with Humans}

\begin{figure*}
    \subfloat[Distribution of subjects across XAI papers that validated with a human evaluation.]{%
        \includegraphics[width=.95\linewidth]{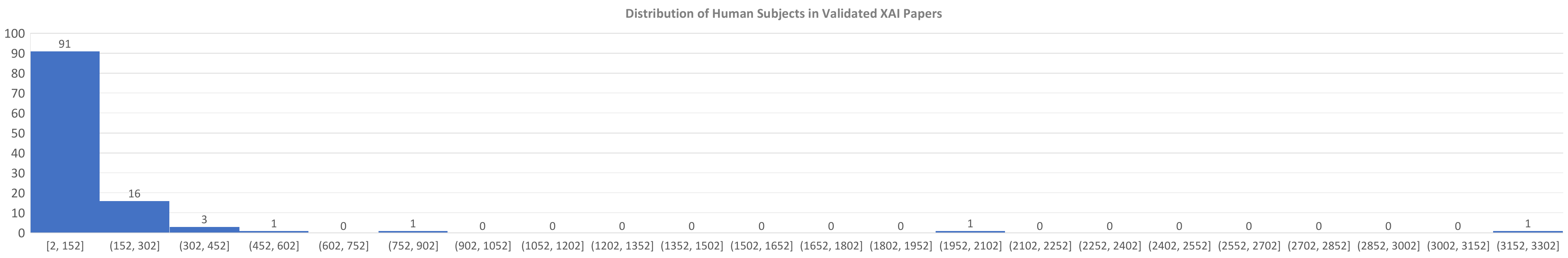}%
        \label{subfig:xai-count}%
    } \\
    \subfloat[Subset of distribution counts, up to N=152 and bin size=5.]{%
        \includegraphics[width=.95\linewidth]{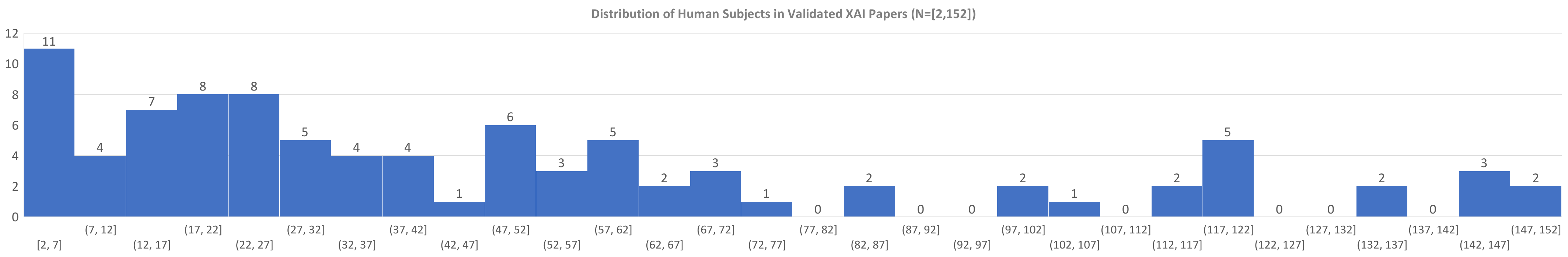}%
        \label{subfig:zoomed-count}%
    }
    \caption{Distribution of counts for human subjects involved in an evaluation, experiment, interview, etc. for validated XAI papers. It is important to note that 13 papers did not report their human subject count and 3 papers approximated their count. In the bottom figure, a subset of distribution counts up to N=152 and a bin size of 5.}
    \Description{Two subfigures depicting histograms, one on the top and one on the bottom. The first histogram shows the distribution of participant counts for the entire range of data, while the bottom histogram represents a zoomed in view from participant count 0 to 152.}
    \label{fig:participant-count}
\end{figure*}

We find that out of the 237 papers that claimed some form of human explainability, only 56\% of them provided an evaluation with humans. 
Of course, this number drops down to 0.7\% when compared to the larger body of XAI work in which empirical evidence is not explicitly mentioned as a keyword in the paper, but for which explainability is a central topic and often a central claim of a new AI technique. 
Even if we extrapolate the fraction of off topic papers in the review set to the entirety of the XAI set, and exclude the same fraction from the overall number of papers under consideration, the percentage of papers with human evaluations as a part of this new ``on-topic'' set is \textit{still} approximately 0.7\%. 

One point worth discussing is the quality of the empirical evaluations in those 128 human-validated papers. In some cases, the participants in the evaluation were never directly discussed. Seen in Figure~\ref{fig:participant-count}, 13 of the papers that involved a validation did not report their human subject count and 3 papers reported an approximate count. Furthermore, in many of the papers, the authors do not disclose the subjects' expertise, workplace, education, or relationship to the authors. We found ourselves asking: \textit{who} are the evaluation subjects?
If we know nothing about these users' backgrounds, or potential motivations to call an XAI technique `good,' one might argue this introduces an unfair bias into the validation.

Furthermore, many of the papers recruited college students as evaluation participants. Prior research has questioned the validity of such experiments~\cite{wells2021explainable} when the end-users of these XAI techniques are intended to be cybersecurity experts, pilots, doctors, and so on. Of course, it can be difficult to assemble domain experts to test methods and tools, especially domain experts who you are not working on the tool with. Perhaps that is why we observed a number of papers only having two or three domain experts as part of the evaluation (Figure~\ref{subfig:zoomed-count}). Regardless, it is difficult to trust generalizable claims about explainability when the participant pool is very small or dissimilar to the target users. 

Another observation was the type of evaluation conducted. Many of the papers’ evaluation criteria consisted of: \textit{``Do you think this is explainable – yes or no?''} or \textit{``Which is more explainable: our explainable method, or this other baseline explanation that is clearly worse (e.g., contains grammatical errors or has logical flaws)?''} It is important that evaluation scenarios are crafted realistically to their intended target use case. Many questions related to the risks, potential failures, or edge-cases remain unanswered when explainable AI techniques are not properly studied with the correct end-user group nor represent a realistic evaluation scenario.    

\subsection{Studies without Humans}
\label{sec:no-humans}

The studies that propose new XAI techniques but lack human testing almost universally show results of algorithmic performance, and at times a few exemplar outputs. While these elements are important and may provide circumstantial evidence, they are insufficient if a central claim is that a method is ``explainable.'' If we think about the production of a possibly-explainable artifact as a signal being sent, it is straightforward for us to realize that it takes confirmation that the signal was received and understood correctly to confirm a claim of explainability. Otherwise, there is no difference between ``explainable'' AI and any other kind of AI, and the many examples of possibly-explainable methods are simply sending signals into a void. 

We previously noted that the field still tends to use terms such as ``explainable,'' ``interpretable,'' ``transparent,'' and so on, interchangeably~\cite{lipton2018mythos}. Rudin~\cite{rudin2019stop} makes the distinction between ``explainable'' as a post-hoc treatment of an otherwise black-box model, ``interpretable'' stemming from model design that is grounded in expert domain understanding. A small number of the papers that received a validation score of 0 did develop their models with the help of domain experts, thereby meeting Rudin’s ``interpretable'' definition, but did not perform any formal evaluation with such experts – neither with the group that helped develop them or with a broader group. Even in these cases, it remains an important step that the evaluation of human understanding---beyond the typical one or two domain experts who helped develop these models---is included. While it is more likely that signals crafted with the help of domain experts are meaningful to their intended audience, a validation that such signals were received and understood is still necessary, as previous work has shown that it is not always the case. For example, domain experts can be vastly overconfident in their understanding of XAI methods that they helped craft~\cite{siu2023stl}. 

For the many other papers with a validation score of 0, there is an often implied or at times explicit claim that an explanatory method is ``intuitively obvious'' due to the inclusion of some property such as the use of natural language or conciseness of explanation. While these properties may seem self-evident, they are neither inherently clear nor universally effective, as even brief texts in one’s native language can often be confusing or indecipherable. These properties are perhaps useful, but not inherently sufficient or necessarily effective~\cite{buccinca2021trust}. 


\section{Future Work}
\label{sec:future}

\subsection{Beyond this Late-Breaking Work}

This late-breaking work intends to shed light on the lackluster empirical evidence in the field of explainable AI. However, we consider this research to be a ``prequel'' to future work that will require refining our literature search, Miller et al.'s scoring criteria~\cite{miller2017explainable}, and our analysis of the resultant literature.

\smallbreak 
\noindent
\textbf{Generalizability.} There are multiple opportunities to build upon our XAI literature review to improve the generalizability of our findings. Future work should examine identifying and analyzing literature not included in our search criteria: review articles, workshop papers, and Non-English papers. Including these works in an analysis would not only help the community gain a deeper understanding of empirical testing in XAI, but would also mitigate the inherent bias of studying English-only works. Our literature search could also be refined to distinguish different types of XAI research, resulting in a potentially smaller set of `All XAI papers' that are deemed more on-topic.  For example, there are cases of XAI work that do not formally evaluate with humans due to their focus on the use of explanations for scientific discovery, or to understand the mechanisms by which some system works. These are typically works in the sciences, and it is true that broader evaluation of human-interpretability is likely unnecessary here if they directly present what principles have been discovered, or in many cases rediscovered, since those principles can be checked directly by the relevant domain experts. We posit these cases form a small minority of the work in the field; however, we would like to consider additional cases that should not be included in our broadest literature search. 

\smallbreak 
\noindent
\textbf{Scoring criteria.} Another important direction of research would be to rework Miller et al.'s scoring criteria~\cite{miller2017explainable} from 2017, specifically the data-driven criteria. Miller et al. rewarded XAI papers with the highest score of a 2 if the paper: (1) explicitly referenced a work in the social sciences \textit{or} in a computer science venue, so long as the paper was not centrally about AI; and (2) incorporated a method from the social sciences in the paper's contributed XAI method. We find that the data-driven criteria is somewhat unclear, and it is difficult to understand which computer science papers `count' versus `do not count.' Should XAI papers that cite other XAI papers---which \textit{do} include empirical evidence---receive a higher score? What if a new XAI method incorporates the results of a prior XAI paper but does not test this new method with human subjects? The validation score we used may also need to be reworked, as the sole presence of an evaluation may not be equivalent to a paper that presents a `good' evaluation (as discussed in Section~\ref{sec:results}, some evaluations did not report a participant count or participant expertise). 

\smallbreak 
\noindent
\textbf{Comprehensive analysis.} It is imperative that a deeper analysis is conducted of our final set of `on-topic' XAI papers. We briefly touched on the number of human subjects across all validated papers (Figure~\ref{fig:participant-count}), but some questions remain. For example, what is the distribution for the \textit{types} of subjects---e.g., experts versus students? Further, what we can learn from the papers that do and do not validate their claims with human studies, but still claim interpretability? What else do these papers have in common, and how else do they differ? Are there shared outcomes we can learn in testing explainability \textit{without} human subjects? Moreover, do papers that validate with human studies tend to be published at a human-centered conference, e.g., ACM CHI, and not at an AI-centered conference, e.g., NeurIPS? The answers to these questions can help identify best practices in XAI testing, implications for AI design, and ultimately pave the way for rigorous human-centered evaluation frameworks. If other researchers would like to tackle these questions, we provide the exact keyword search used for our literature search in Table~\ref{tab:scopus-search}, which can be employed to replicate our findings.

\subsection{Call to Action}
Like claims of AI accuracy, claims of AI explainability must rest on empirical evidence. Contemporary artificial intelligence is an empirical discipline that rests on evidence of accuracy. However, our results show that work in contemporary explainable artificial intelligence almost never provides empirical evidence of its own. 

While general heuristics such as conciseness or the use of domain-specific features are a reasonable place to start, we must not only rely on these heuristics. Empirical human testing is necessarily the gold standard for explainability claims. A pharmaceutical company would never distribute a newly developed drug without extensive clinical trials, even if they truly believed in its efficacy from biological principles, evidence collected in simulation, mouse trials, and so on. Why can XAI models, which may impact high-risk decision-making, be released untested with their human users?  

Common objections to empirical evaluations are that humans are biased, or it is difficult to assemble a representative sample of humans. We note that the same can be said of every commonly-used machine learning benchmark, from ImageNet~\cite{deng2009imagenet}, to IMBD movie reviews~\cite{maas2011learning}, to medical question-and-answer~\cite{jin2021disease}. All contain human biases from their generation, curation, cultural context, or other factors. They may or may not match the intended application of the system being tested and took extensive time to collect. Nonetheless, they are used. The issue of limitations in the human testing population in interpretability testing is analogous to limitations in the dataset being used in accuracy testing and is not an unknown to the AI community.  

The methods to test XAI clearly exist~\cite{doshi2017towards}---our literature review shows that there are researchers using them and publishing about them. Undoubtedly there are challenges in moving from testing AI solely with static datasets into testing with humans, but they are not insurmountable. It is crucial that the XAI community take a critical look at the state of the field and determine whether the current trajectory and (lack of) standards is acceptable or not. As researchers, it seems unacceptable to produce copious volumes of ``explainable'' methodologies that have never been tested with the intended end-users. As reviewers, we are responsible for pushing back on research that claims explainability but provides little to no evidence. We hope the findings presented in this work challenge each other's viewpoints.

\section{Conclusion}

This late-breaking work highlights a critical gap in explainable AI (XAI) research: the overwhelming majority of papers that claim explainability fail to provide empirical evidence through human studies. By conducting a comprehensive literature search of 18,254 papers in collaboration with a professional librarian, we identified 253 papers that explicitly referenced human-related terms, such as human studies, user experiments, or end-user feedback. Among these, only 128 included some form of a human evaluations, revealing that 0.7\% of XAI literature empirically validates claims of explainability. These findings highlight an urgent need for greater rigor in XAI research, particularly in substantiating claims about human interpretability and usability. Without empirical evidence, it is impossible to assess whether explainability methods truly address the needs of end-users. To foster transparency and reproducibility, we share our methodology and keyword search criteria, encouraging researchers to build on our work and prioritize human-centered evaluation in future explainable AI work.

\begin{acks}
DISTRIBUTION STATEMENT A. Approved for public release. Distribution is unlimited. This material is based upon work supported by the Under Secretary of Defense for Research and Engineering under Air Force Contract No. FA8702-15-D-0001. Any opinions, findings, conclusions or recommendations expressed in this material are those of the author(s) and do not necessarily reflect the views of the Under Secretary of Defense for Research and Engineering.
\end{acks}

\bibliographystyle{acm-contents/ACM-Reference-Format}
\balance
\bibliography{xai}

\end{document}